\documentclass[10pt,letterpaper]{article}
\usepackage[top=0.85in,left=1.25in,footskip=0.75in,marginparwidth=2in]{geometry}

\usepackage[utf8]{inputenc}

\usepackage{nameref,hyperref}

\usepackage[right]{lineno}

\usepackage{microtype}
\DisableLigatures[f]{encoding = *, family = * }

\raggedright
\usepackage{changepage}

\usepackage[aboveskip=1pt,labelfont=bf,labelsep=period,singlelinecheck=off]{caption}

\makeatletter
\renewcommand{\@biblabel}[1]{\quad#1.}
\makeatother

\usepackage{lastpage,fancyhdr,graphicx}
\usepackage{epstopdf}
\usepackage{color}

\definecolor{Gray}{gray}{.25}

\usepackage{graphicx}

\usepackage{sidecap}

\usepackage{wrapfig}
\usepackage[pscoord]{eso-pic}
\usepackage[fulladjust]{marginnote}
\reversemarginpar

\begin{document}
\vspace*{0.35in}

\begin{flushleft}
{\Large
\textbf\newline{Characterizing the deep uncertainties surrounding coastal flood hazard projections: A case study for Norfolk, VA}
}
\newline
\\
K. L. Ruckert\textsuperscript{1},
V. Srikrishnan\textsuperscript{1},
K. Keller\textsuperscript{1,2,*}
\\
\bigskip
\bf{1} Earth and Environmental Systems Institute, The Pennsylvania State University, University Park, Pennsylvania, USA
\\
\bf{2} Department of Geosciences, The Pennsylvania State University, University Park, Pennsylvania, USA
\\
\bigskip
* klaus@psu.edu

\end{flushleft}

\section*{Abstract}
Coastal planners and decision makers design risk management strategies based on hazard projections. However, projections can differ drastically. What causes this divergence and which projection(s) should a decision maker adopt to create plans and adaptation efforts for improving coastal resiliency? Using Norfolk, Virginia, as a case study, we start to address these questions by characterizing and quantifying the drivers of differences between published sea-level rise and storm surge projections, and how these differences can impact efforts to improve coastal resilience. We find that assumptions about the complex behavior of ice sheets are the primary drivers of flood hazard diversity. Adopting a single hazard projection neglects key uncertainties and can lead to overconfident projections and downwards biased hazard estimates. These results highlight key avenues to improve the usefulness of hazard projections to inform decision-making such as (i) representing complex ice sheet behavior, (ii) covering decision-relevant timescales beyond this century, (iii) resolving storm surges with a low chance of occurring (e.g., a 0.2\% chance per year), (iv) considering that storm surge projections may deviate from the historical record, and (v) communicating the considerable deep uncertainty.

\section*{Introduction}
\paragraph{}
Coastal flood hazards are increasing in many regions around the world \cite{Kopp2014-ri, Intergovernmental_Panel_on_Climate_Change2013-sl}. Decision makers  are designing strategies to manage the resulting risks \cite{National_Research_Council1987-oj, US_Army_Corps_of_Engineers2011-fm, US_Army_Corps_of_Engineers2013-oh, US_Army_Corps_of_Engineers2014-jt, Norfolk-2015-rs, Hall2016-fh, Coastal_Protection_and_Restoration_Authority_of_Louisiana2017-sn, US_Army_Corps_of_Engineers2018-nv}. The design of such flood risk management strategies can hinge critically on flood hazard projections \cite{Keller2015-uk, Sriver2018-fw}. Decision makers face a potentially confusing array of flood hazard projections. These projections include scenarios without formal probabilistic statements (e.g., \cite{US_Army_Corps_of_Engineers2011-fm, US_Army_Corps_of_Engineers2013-oh, US_Army_Corps_of_Engineers2014-jt, Parris2012-oz, Hall2016-fh}), single probability density functions (e.g., \cite{Sweet2017-uh}), and probabilistic scenarios (i.e., multiple probability density functions conditional on model assumptions) (e.g., \cite{Kopp2014-ri, Kopp2017-fz, Wong2017-kn, Rasmussen2018-uy}).  Furthermore, the projections differ in crucial assumptions, for example, about the potential non-stationarity of storm surges and/ or about future potential abrupt changes in ice-sheet dynamics \cite{Kopp2017-fz, Wong2017-kn,Wong-2018}. Here, we synthesize and analyze published flood hazard projections.  We hope that this synthesis can help to improve the understanding of what drives the apparent diversity of coastal flood hazard projections and, in turn, can help to improve the design of flood risk management strategies.

\paragraph{}
The synthesis and analysis of flood hazard projections is an area of active research with a rich body of excellent previous work. Relevant examples include the U.S. Army Corps of Engineers~\cite{US_Army_Corps_of_Engineers2011-fm, US_Army_Corps_of_Engineers2013-oh, US_Army_Corps_of_Engineers2014-jt}, Tebaldi et al.~\cite{Tebaldi2012-ou}, Parris et al.~\cite{Parris2012-oz}, Zervas~\cite{Zervas2013-sv}, Kopp et al.~\cite{Kopp2014-ri}, Hall et al.~\cite{Hall2016-fh}, Kopp et al.~\cite{Kopp2017-fz}, Sweet et al.~\cite{Sweet2017-uh}, Wong and Keller~\cite{Wong2017-kn}, Rasmussen et al.~\cite{Rasmussen2018-uy}, and Wong~\cite{Wong-2018}. These studies have broken important new ground, but they are hard to compare due to differences in underlying assumptions and projection structure. Further, decision makers assess community vulnerability, and design and implement flood risk management strategies on a local to regional scale. Hence, we focus on a case study for the Sewell’s Point tide gauge in Norfolk, Virginia, USA, to be relevant to local-scale preparedness planning and risk management. We choose the city of Norfolk as a case study because it is prone to impacts from sea-level rise (SLR), nuisance flooding from high tides, heavy rainfall, and as well as tropical and extra-tropical storms. Additionally, it is the location of an active U.S. Navy base (Naval Station Norfolk). Although we restrict this study to the city of Norfolk, we analyze differences in published flood hazard projections that can be localized to other coastal cities following our methodology. As a result, this study can provide useful insights to the broader community interested in local coastal protection.

We consider multiple future SLR scenarios and characterizations of storm surge generated from different approaches to represent a range of choices in the coastal assessment, planning, and decision-making process. We expand upon the current state-of-the-art by assessing the differences, the potential consequences of these differences, and addressing the results in a local coastal protection context. We compare eight studies of SLR \cite{US_Army_Corps_of_Engineers2011-fm, US_Army_Corps_of_Engineers2013-oh, US_Army_Corps_of_Engineers2014-jt, Parris2012-oz, Kopp2014-ri, Hall2016-fh, Kopp2017-fz, Sweet2017-uh, Wong2017-kn, Rasmussen2018-uy}, four studies of storm surge \cite{US_Army_Corps_of_Engineers2014-jt, Tebaldi2012-ou, Zervas2013-sv, Wong-2018}, and one storm surge analysis that is new to this work. We choose to compare these studies for three reasons: 1) they depict knowledge gained over a decade of research, 2) they integrate global SLR scenarios with regional factors, and 3) the values intend to support stakeholder groups and communities in regard to coastal preparedness planning and risk management. The overall goal of this work is to evaluate the current scientific knowledge to identify and highlight current limitations and community needs that can support real coastal preparedness planning and risk management processes.

\section*{Results}
\subsection*{Sea-level rise projections}
\begin{figure}[ht]
\centering
\includegraphics{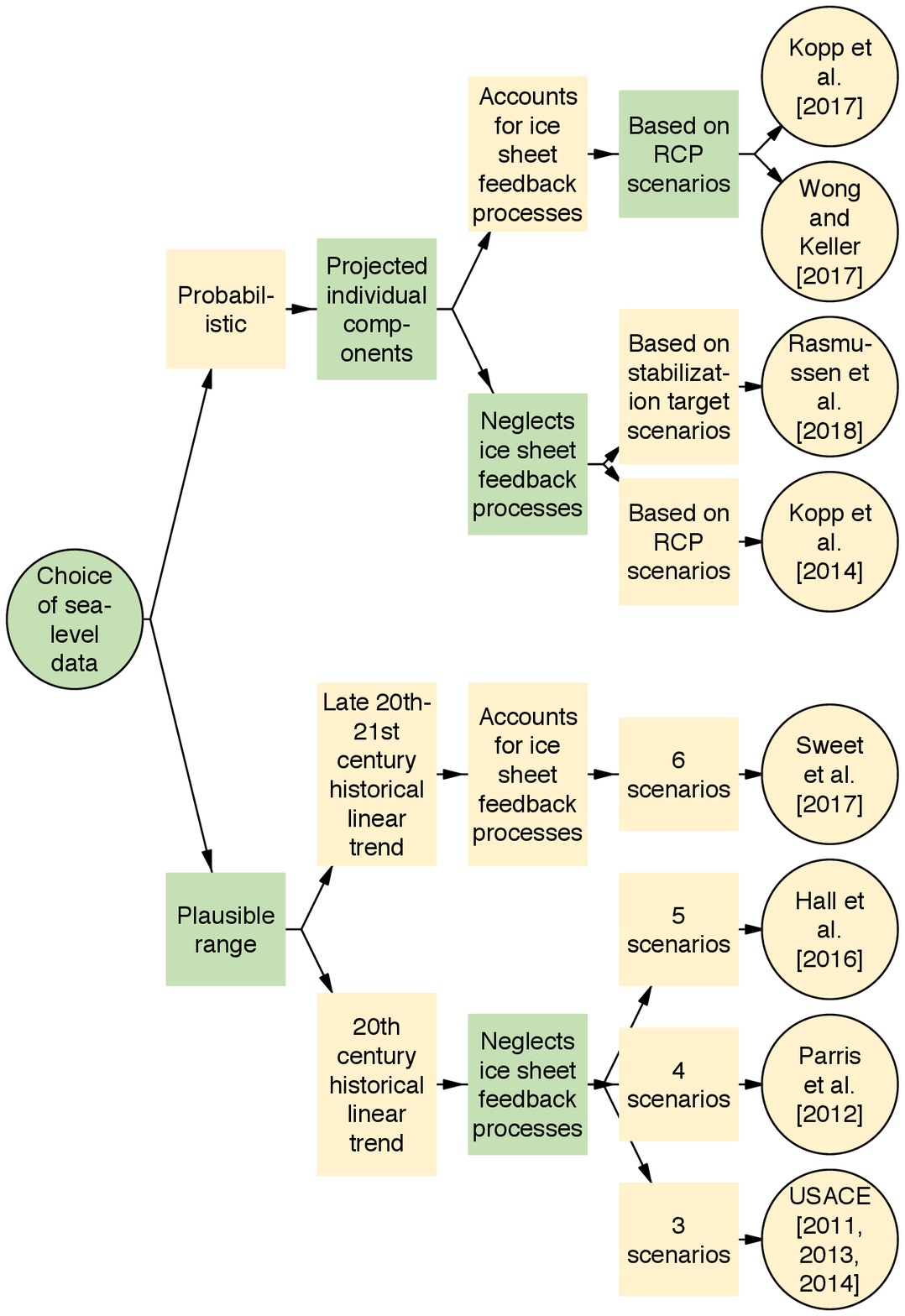}
\caption{Classification diagram visualizing the differences between the considered sea-level projections.}
\label{figone}
\end{figure}

\paragraph{}
The SLR scenarios evaluated here follow two different methods (Figure~\ref{figone}). The first method provides probabilistic projections of individual components of SLR for representative concentration pathways (RCP) \cite{Moss-2010} or target temperature stabilization scenarios, which are then downscaled to the local level \cite{Kopp2014-ri, Kopp2017-fz, Wong2017-kn, Rasmussen2018-uy}. The other considered studies follow the method of providing scenarios that describe plausible conditions across a broad range representing the scientific knowledge at the time of report development \cite{US_Army_Corps_of_Engineers2011-fm, US_Army_Corps_of_Engineers2013-oh, US_Army_Corps_of_Engineers2014-jt, Parris2012-oz, Hall2016-fh, Sweet2017-uh}.

\subsection*{Probabilistic projections}
\paragraph{}
Kopp et al.~\cite{Kopp2014-ri}, Kopp et al.~\cite{Kopp2017-fz}, Wong and Keller~\cite{Wong2017-kn}, and Rasmussen et al.~\cite{Rasmussen2018-uy} are examples of probabilistic projections of SLR, which can be localized to the Sewell's Point tide gauge. These studies are all based on the framework from Kopp et al.~\cite{Kopp2014-ri} with the exception of Wong and Keller~\cite{Wong2017-kn}. As such, understanding the method and framework behind Kopp et al.~\cite{Kopp2014-ri} is of great importance to flood risk management strategies. Although Kopp et al.~\cite{Kopp2014-ri} provides a valuable step forward in developing local probabilistic projections based on individual SLR components, it faces limitations with respect to the projection of the complex behaviors of the ice sheets, as well as the consideration of stabilization targets. Relevant studies that expand on these aspects include Kopp et al.~\cite{Kopp2017-fz}, Wong and Keller~\cite{Wong2017-kn}, and Rasmussen et al.~\cite{Rasmussen2018-uy}.

Kopp et al.~\cite{Kopp2014-ri} defines a set of probabilistic global and local SLR projections constructed with RCP scenarios by modeling individual processes that contribute to local SLR. The individual components include oceanic processes, ice sheet melt, glacier and ice cap surface mass balance, land-water storage, and long-term, local, non-climatic sea-level change. To calculate global sea-level probability distributions, Kopp et al.~\cite{Kopp2014-ri} employs 10,000 Latin hypercube samples from cumulative SLR contributions. Kopp et al.~\cite{Kopp2014-ri} then localizes these projections (i.e., at tide gauge locations) by applying sea-level fingerprints \cite{Mitrovica2011-hm}.

Kopp et al.~\cite{Kopp2017-fz} employs the same framework as Kopp et al.~\cite{Kopp2014-ri}, except that it replaces the AIS projections with those based on DeConto and Pollard~\cite{DeConto2016-dp}. The original AIS projections used in Kopp et al.~\cite{Kopp2014-ri} account for simple assumptions of constant acceleration that underlies expert-judgement-based projections. Instead, the new approach influences buttressing ice shelves and hence accounts for marine ice cliff instability and hydrofracturing (for more details see \cite{DeConto2016-dp}). However, the ensemble of AIS projections were developed using a simplified approach of sampling key physical parameters from a set of values and integrating paleo-observations with a pass/ fail test rather than producing a probability distribution. Because Kopp et al.~\cite{Kopp2017-fz} directly uses these AIS projections, the resulting SLR projections are potentially more conservative with respect to low probability AIS projections.

Wong and Keller~\cite{Wong2017-kn} employ two sets of simulated sea-level scenarios. One scenario assumes that there is no contribution of fast Antarctic ice sheet dynamics (e.g., ice cliff instability and hydrofracturing), while the other assumes that the fast dynamics is triggered. Wong and Keller~\cite{Wong2017-kn} emulate fast dynamics with a simplified approach that assumes a constant rate of disintegration once a critical temperature threshold is passed.  For this study, we differentiate the scenarios as Wong and Keller~\cite{Wong2017-kn} FD and Wong and Keller~\cite{Wong2017-kn} no FD, respectively assuming fast dynamics is triggered and assuming fast dynamics is not triggered. Specifically, we use the results based on prior gamma distributions for the parameters that control the uncertain rate of disintegration and the threshold temperature that triggers fast dynamical disintegration.

Wong and Keller~\cite{Wong2017-kn} use a simple mechanistically motivated emulator to project coastal flooding hazards (BRICK: Building Blocks for Relevant Ice and Climate Knowledge model v0.2) \cite{Wong2017-le}. The BRICK model simulates global mean surface temperature, ocean heat uptake, thermal expansion, changes in land-water storage, and ice melt from the Antarctic ice sheet, Greenland ice sheet, and glaciers and ice caps. Wong and Keller~\cite{Wong2017-kn} calibrate this model to observational records (paleoclimate and instrumental data) using a Bayesian approach. We use this  model to extend and derive localize sea-level projections to the Sewell’s Point tide gauge (see details in Methods).

Rasmussen et al.~\cite{Rasmussen2018-uy} models local relative sea level using the Kopp et al.~\cite{Kopp2014-ri} framework (as described above). The authors construct alternative ensembles that meet global mean surface temperature (GMST; relative to 2000) stabilization target scenarios. These scenarios stabilize warming at 1.5, 2.0, and 2.5 $^{\circ}$C above pre-industrial levels, coinciding with targets identified in the Paris Agreement \cite{Unfccc2015-ct, Unfccc2015-ly, Rasmussen2018-uy}. To ensure each scenario meets the stabilization target criteria, only the models that have a 21st century increase in GMST (extrapolated from the 2070 to 2090 trend) of 1.5, 2.0, and 2.5 $^{\circ}$C ($\pm$ 0.25 $^{\circ}$C) are used to create the ensembles. Scenarios beyond 2100 are ensembles that undershoot the target temperature with the exception of the 2.5 $^{\circ}$C scenario.

\subsubsection*{Plausible scenarios}
\paragraph{}
The U.S. Army Corps of Engineers~\cite{US_Army_Corps_of_Engineers2011-fm, US_Army_Corps_of_Engineers2013-oh, US_Army_Corps_of_Engineers2014-jt}, Parris et al.~\cite{Parris2012-oz}, Hall et al.~\cite{Hall2016-fh}, and Sweet et al.~\cite{Sweet2017-uh} all adopt an approach of providing a broad range of future conditions based on published studies. These studies linearly extract the historical tide gauge rate for the lowest scenario and use a global mean SLR model to represent non-linear scenarios. Specifically, they use a quadratic global mean SLR model \cite{National_Research_Council1987-oj} in time (modified to begin in the year 2000 and to project in feet) for eustatic SLR.

The U.S. Army Corps of Engineers~\cite{US_Army_Corps_of_Engineers2011-fm, US_Army_Corps_of_Engineers2013-oh, US_Army_Corps_of_Engineers2014-jt} studies provide three scenarios of relative SLR: 1) a low scenario based on a linear extrapolation of the historical tide gauge rate, 2) an intermediate scenario, and 3) a high scenario (details on downscaling are provided in Methods, Supplementary Table 1).

Parris et al.~\cite{Parris2012-oz} expands on the research conducted in the U.S. Army Corps of Engineers'~\cite{US_Army_Corps_of_Engineers2011-fm, US_Army_Corps_of_Engineers2013-oh, US_Army_Corps_of_Engineers2014-jt} by adding a fourth scenario and modifying scenarios based on scientific research of ocean warming and ice sheet loss (Supplementary Table 1). The highest scenario derives from the ocean warming estimates in the Meehl et al.~\cite{Intergovernmental_Panel_on_Climate_Change2007-sb} global SLR projections along with the maximum estimates of glacier and ice sheet loss in Pfeffer et al.~\cite{Pfeffer2008-ow}. The intermediate-high scenario is derived from the average of high end, semi-empirical, global SLR projections \cite{Horton2008-fr, Vermeer2009-tz, Grinsted2010-lc, Jevrejeva2010-sp}. The intermediate-low scenario is based on the B1 emissions scenario global SLR projection from Meehl et al.~\cite{Intergovernmental_Panel_on_Climate_Change2007-sb} and lastly, the lowest scenario is a linear extrapolation of the historical SLR rate from 20th century tide gauge records \cite{Church2011-po} (details on downscaling in Methods, Supplementary Table 1).

Hall et al.~\cite{Hall2016-fh} uses the same low and high scenario as Parris et al.~\cite{Parris2012-oz}, but proposes intermediate scenarios that are 0.5 m (~1.6 ft) increment subdivisions (Supplementary Table 1). The use of equally proportional subdivisions is chosen due to the imprecise nature of estimating future SLR, associated uncertainties, and the fact that this information is used for vulnerability, impact, and risk management purposes. Unfortunately, downscaled projections calculated by Monte Carlo resampling of fingerprints from Perrette et al.~\cite{Perrette2013-xf} and Kopp et al.~\cite{Kopp2014-ri} are not publicly available due to the sensitive nature of the data. Instead we downscale projections following an approach in the U.S. Army Corps of Engineers~\cite{US_Army_Corps_of_Engineers2011-fm, US_Army_Corps_of_Engineers2013-oh, US_Army_Corps_of_Engineers2014-jt} (details in Methods).

Sweet et al.~\cite{Sweet2017-uh} provides an update of scenarios based on the National Research Council~\cite{National_Research_Council1987-oj} global mean SLR model (Supplementary Table 1). These scenarios include the same intermediate scenarios as in Hall et al.~\cite{Hall2016-fh}, as well as an additional extreme case scenario and an updated low scenario. The upward revision of the low scenario is based on the 3 mm/yr global mean sea-level rate measured from tide gauges and satellite altimeters over the past quarter century \cite{Nerem2010-jr, Church2011-po, Boening2012-yi, Fasullo2013-bx, Cazenave2014-oq, Hay2015-ee}. Sweet et al.~\cite{Sweet2017-uh} adds a worst-case scenario to account for potential acceleration of ice sheet mass loss from physical feedbacks \cite{DeConto2016-dp} and the growing number of studies with global mean sea level that exceeds 6.6 ft by 2100 \cite{Sriver2012-vu, Bamber2013-kg, Miller2013-no, Rohling2013-zv, Jevrejeva2014-pe, Kopp2014-ri, Grinsted2015-in, Jackson2016-bd}.

To project global and regional SLR, Sweet et al.~\cite{Sweet2017-uh} follows the Kopp et al.~\cite{Kopp2014-ri} framework. Specifically, Sweet et al.~\cite{Sweet2017-uh} drives global and regional projections with RCP2.6, RCP4.5, and RCP8.5 and produces 20,000 Monte Carlo samples for each emissions scenario. Regional sea levels (relative to the year 2000) are projected on a 1-degree grid accounting for locations of the tide gauges. At each grid cell, the SLR scenarios are adjusted to account for shifts in oceanographic factors (e.g., circulation patterns), glacial isostatic adjustment from the melting of land-based ice, and non-climatic factors. Sweet et al.~\cite{Sweet2017-uh} then combines the results from each emissions scenario and divides them into subsets according to the six scenarios. These subset distributions are not equal in sample size. 

\subsection*{Storm surge projections}
\paragraph{}
We compare stationary (i.e., not time varying) storm surge projections from four studies \cite{Tebaldi2012-ou, Zervas2013-sv, US_Army_Corps_of_Engineers2014-jt, Wong-2018} to historical observations and projections from an alternative model discussed below (details in Methods). Additionally, we compare stationary storm surge values to non-stationary values in the year 2065. These values are available to decision makers for the Sewell's Point tide gauge location, are relative to the current NOAA national tidal datum epoch (NTDE; 1983-2001) local mean sea level (MSL), and are compared to historical observations \cite{National_Oceanic_and_Atmospheric_Administration2013-ly}.

Zervas~\cite{Zervas2013-sv} analyzes monthly mean highest water levels over a period from 1927-2010. In order to remove the longer-term signal, Zervas~\cite{Zervas2013-sv} linearly detrends the data by removing the mean sea-level trend (based on data up to 2006), which is relative to the NTDE midpoint. These detrended monthly extremes are used to obtain the annual block maximum (the maximum observation in each year) if a year has four or more months of data. If a year has less than four months of data, then no annual block maxima is estimated for that year. Zervas~\cite{Zervas2013-sv} fits the annual block maxima to a Generalized Extreme Value (GEV) distribution using the extRemes R package \cite{Gilleland2013-rj, R_Core_Team2016-yv} for estimation of the location, scale, and shape parameters. Using the maximum likelihood estimate of the GEV parameters and a range of exceedance probabilities, Zervas~\cite{Zervas2013-sv} approximates flood return levels with a 95\% confidence interval.

The U.S. Army Corps of Engineers~\cite{US_Army_Corps_of_Engineers2014-jt} study uses the same historic monthly extreme water level values as Zervas~\cite{Zervas2013-sv}, but analyzes a shorter time period from 1927 to 2007. Instead of following the GEV approach laid out in Zervas~\cite{Zervas2013-sv}, the U.S. Army Corps of Engineers~\cite{US_Army_Corps_of_Engineers2014-jt} study follows a percentile statistical function~\cite{Kriebel-2013} and only presents return periods that are within the time frame of the data record. For instance, they do not present the 100-yr return period for Sewell's Point tide gauge because the data record is less than 100 years in length.

Tebaldi et al.~\cite{Tebaldi2012-ou} uses a combination of hourly (1979-2008) and monthly (1959-2008) data. Assuming the long-term trends in local sea level are linear, Tebaldi et al.~\cite{Tebaldi2012-ou} detrends the hourly data using a linear model fit to the monthly data. These detrended hourly values are used to compute the daily maxima and to perform a peak-over-threshold (POT) analysis. Tebaldi et al.~\cite{Tebaldi2012-ou} performs a POT analysis by selecting a threshold corresponding to the 99th percentile and identifying daily values exceeding that threshold. To avoid counting a storm twice, Tebaldi et al.~\cite{Tebaldi2012-ou} uses a 1-day declustering timescale identifying the maximum value among consecutive extremes. The exceedance values identified in the POT analysis are fit to a Generalized Pareto distribution (GPD) for parameter estimation. Using the maximum likelihood estimate of the GPD parameters, Tebaldi et al.~\cite{Tebaldi2012-ou} computes flood return levels and return periods with a 95\% confidence interval.

Wong~\cite{Wong-2018} analyzes 86 years (1928-2013) of hourly data from the tide gauge to generate storm surge projections. First, Wong~\cite{Wong-2018} detrends the data by subtracting a moving window 1-year average and calculates the daily maximum sea levels with the detrended data. Like the analysis in Tebaldi et al.~\cite{Tebaldi2012-ou}, Wong~\cite{Wong-2018} uses the 99th percentile as the threshold for extreme events. However, Wong~\cite{Wong-2018} differs by using a declustering timescale of 3 days to identify the maximum value among consecutive extremes. The exceedance values are then fit to a GPD model for parameter estimation using a Bayesian calibration approach with an adaptive Metropolis Hastings algorithm, where non-stationarity is incorporated into the parameters. Non-stationarity is incorporated using several covariates: time, sea level, global mean temperature, the North Atlantic Oscillation (NAO) index, and a combination of all the covariates generated by applying Bayesian mode 

\subsection*{Differences in flood hazard projections}
\begin{figure}[ht]
\centering
\includegraphics{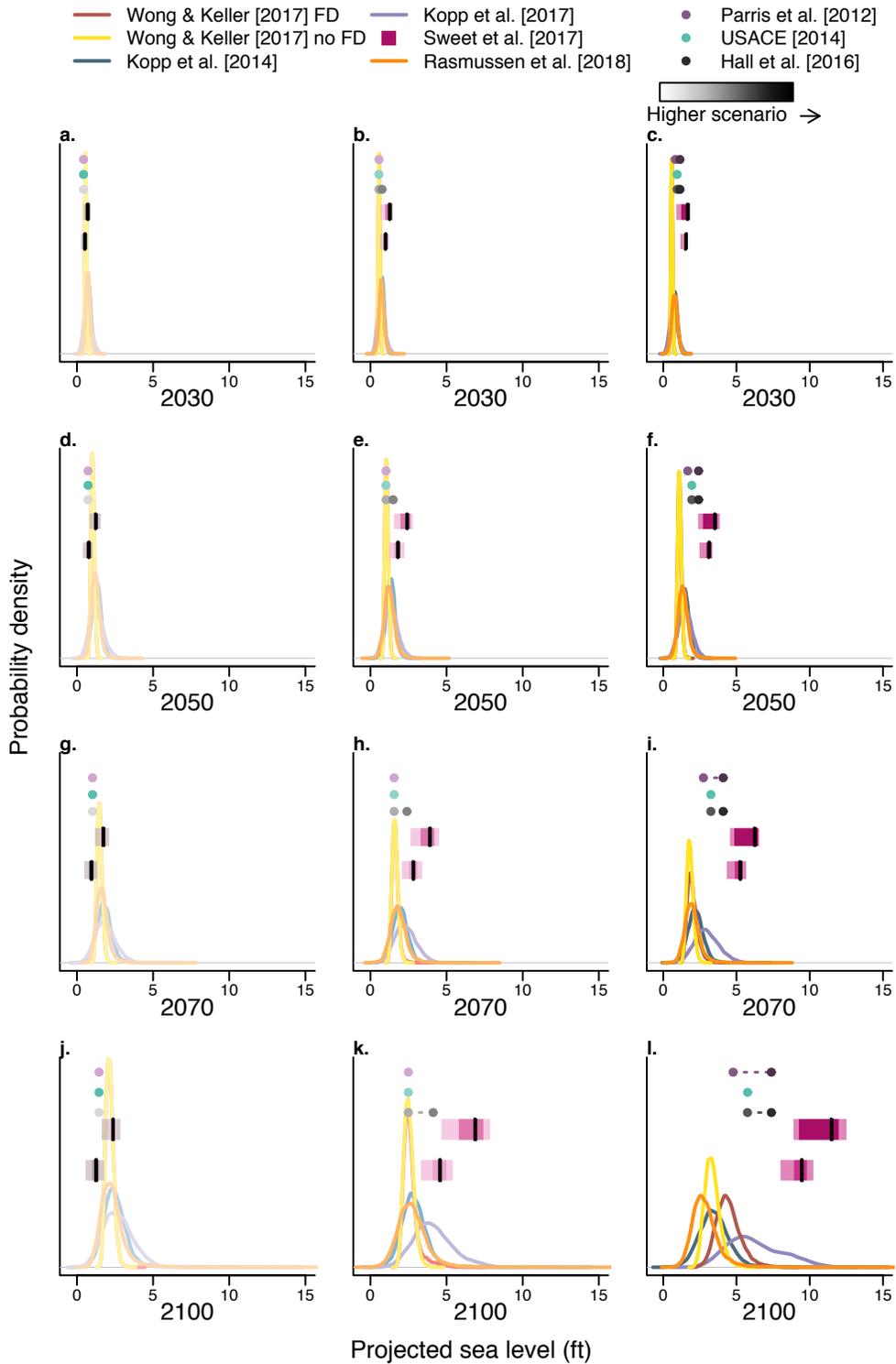}
\caption{Comparison of localized sea-level rise projections for Norfolk, VA. Panels from left to right are projected with higher emissions scenarios (increasing in shade) and panels from top to bottom increase in time from 2030 to 2100. "FD" refers to fast Antarctic ice sheet dynamics and blocks depict the 5, 25, 50, 75, and 95\% quantiles for Sweet et al.~\cite{Sweet2017-uh}.}
\label{figtwo}
\end{figure}

\paragraph{}
We characterize the differences in published SLR projections for the city of Norfolk, VA. The localized SLR projections become less certain and increasingly diverge as time goes on (Figure~\ref{figtwo}). The divergence between projections becomes more apparent with those based on high emissions scenarios and those accounting for uncertainty (Figure~\ref{figtwo}C, F, I, L). It is especially noticeable in the year 2100 when comparing projection modes (Figure~\ref{figtwo}L). These differences can not only be traced back to the assumptions made, but they can also impact coastal preparedness planning.

\subsection*{Accounting for ice sheet feedback processes increases sea-level rise projections}
\begin{figure}[ht]
\centering
\includegraphics{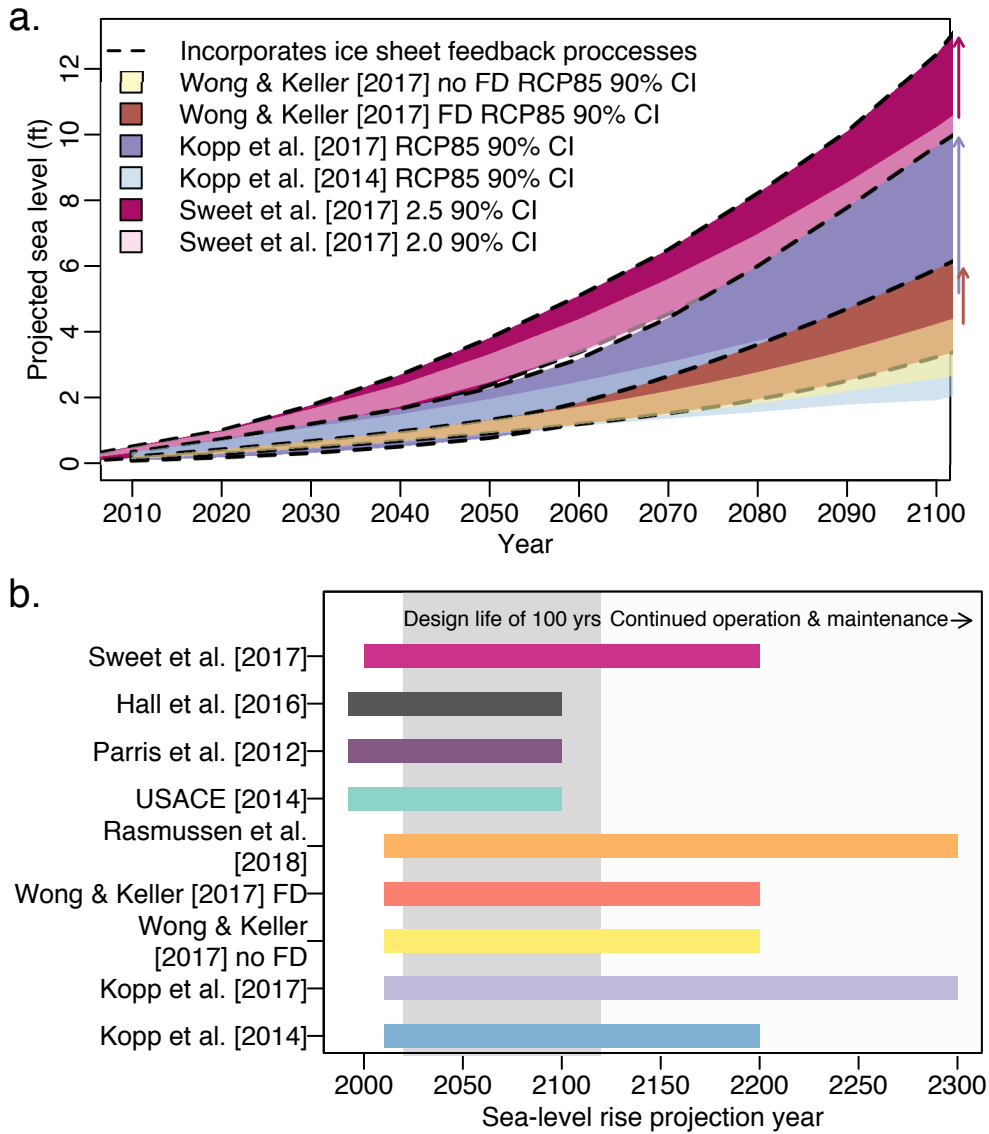}
\caption{Comparison of different assumptions between sea-level projections. Panel A) compares 90\% credible/ confidence intervals of high scenario projections that differ with respect to ice sheet assumptions. Opaque polygons with a dashed border represent projections accounting for ice sheet feedbacks. The arrows highlight the divergence between projections from the same model that differ by ice sheet assumptions. Panel B) compares the timescale of projections. The gray blocks symbolize the potential design life of infrastructure built in 2020.}
\label{figthree}
\end{figure}

\paragraph{}
In the 21st century, local SLR projections differ depending on the assumptions with respect to ice sheet processes (Figure~\ref{figtwo} and Figure~\ref{figthree}A). Studies that incorporate ice sheet feedback processes project estimates of SLR that are greater than those that do not incorporate these processes. Compare, for example, the 95th percentile in 2100 of Wong and Keller~\cite{Wong2017-kn} FD to Wong and Keller~\cite{Wong2017-kn} no FD.  Additionally, compare the 95th percentile in 2100 of Kopp et al.~\cite{Kopp2017-fz} to Kopp et al.~\cite{Kopp2014-ri} and also Sweet et al.~\cite{Sweet2017-uh} (2.5m) to Sweet et al.~\cite{Sweet2017-uh} (2.0m) (Figure~\ref{figthree}A). In all three cases, the projections incorporating ice sheet feedback processes project higher SLR by the year 2100 with an increase of roughly 1.7 to 4.5 ft (comparing the 95th percentiles). Moreover, the divergence between studies with and without feedback processes grows over time with an increase in acceleration in the later half of the century. These projections indicate that ice sheets play a small role in the projected SLR contributions during the first half of the century and a larger role in the second half. This change occurs when anthropogenic greenhouse gas emissions (particularly high emissions scenarios) trigger ice-cliff and ice shelf feedback processes in the Antarctic ice sheet. The results in DeConto and Pollard~\cite{DeConto2016-dp} suggest that the role of ice sheets in SLR contributions will continue to grow in the centuries following 2100. Hence, it is important to account for individual components comprising SLR in addition to ice sheet feedback processes because components interact on different timescales.

\subsection*{Length of projection time can impact long-term adaptation strategies}
\paragraph{}
Of the evaluated studies, half focus on timescales of 110 years or less (excluding the fact that we extend the projection of Wong and Keller~\cite{Wong2017-kn} to 2200)(Figure~\ref{figthree}B). This lack of information can pose problems for the design of coastal adaptation strategies. For instance, the U.S. Army Corps of Engineers typically designs projects to last for 20 to 100 years \cite{US_Army_Corps_of_Engineers2014-jt}. Yet, infrastructure often extends past its original design life due to continued operation and maintenance \cite{US_Army_Corps_of_Engineers2014-jt}. Consider for example, a project designed in the year 2020 with a design life of 100 years. This project could extend well past the year 2120 and would require SLR information for at least 20 years past 2100 for decision making (Figure~\ref{figthree}B). Hence, providing information about SLR beyond 2100 has the potential to improve the robustness and resilience of infrastructure as well as long-term coastal adaptation strategies. However, it is important to be cautious about long-term projections if these projections are based on models with simplified ice dynamics.

\subsection*{Lack of information about storm surge analysis can lead to surprises}
\begin{figure}[ht]
\centering
\includegraphics{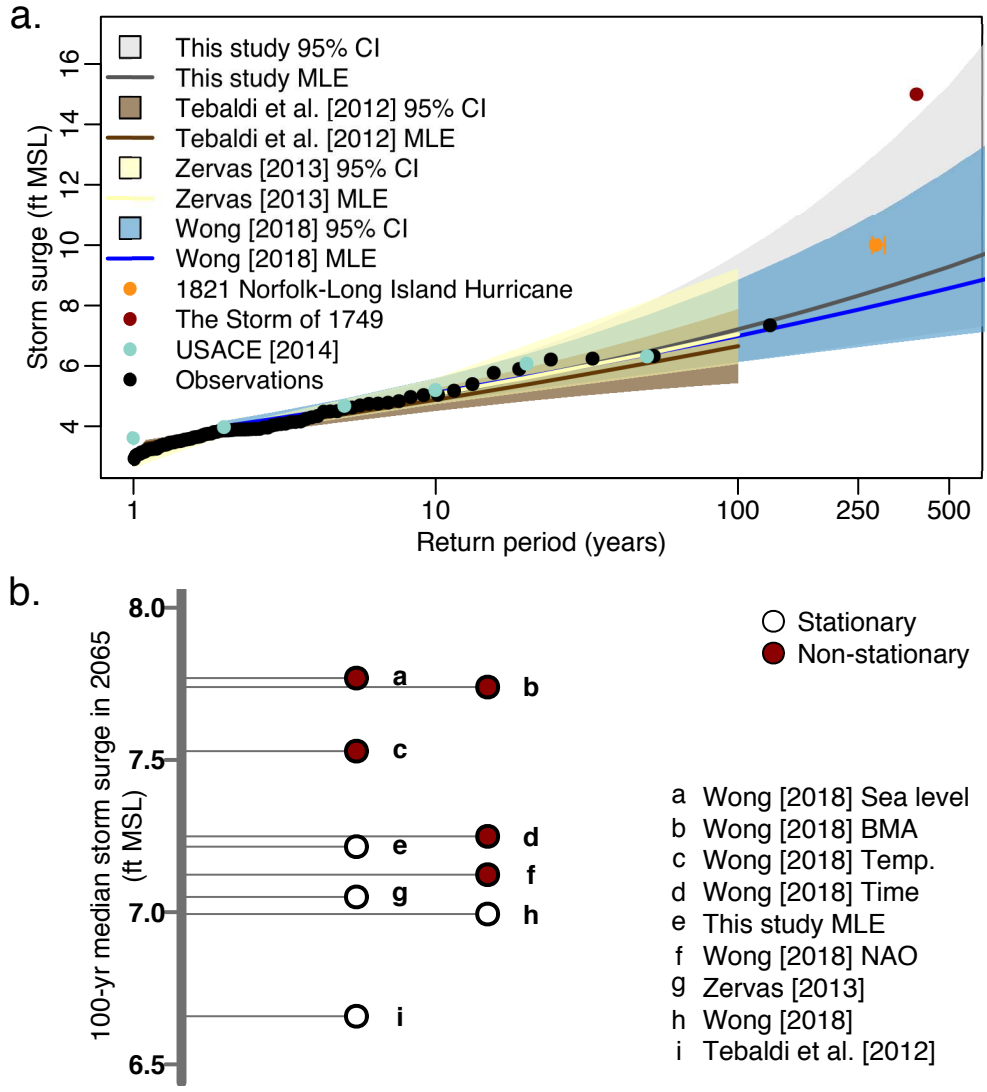}
\caption{Comparison of different assumptions between storm surge projections. Panel A) compares stationary storm surge levels with their associated return periods (inverse flood probability). Panel B) compares the median 100-yr storm surge values based on stationary models to those based on non-stationary models in the year 2065.}
\label{figfour}
\end{figure}

\paragraph{}
Areas prone to storm surge like Norfolk also require information about the frequency of extreme water-level events when defining coastal vulnerability. In particular, analyses resolving these extreme events, especially long return periods, require long records of data (70+ years) to stabilize estimates \cite{Wong2017-ns}. Our results are consistent with this conclusion. For example, the projections based on a 49 year record in Tebaldi et al.~\cite{Tebaldi2012-ou} produce a low bias for long return periods in comparison to the observations, our model, and other studies, which are all based on records of 80 to 90 years in length \cite{Zervas2013-sv,US_Army_Corps_of_Engineers2014-jt, Wong-2018} (Figure~\ref{figfour}A).

Based on historical events, additionally it is important to resolve and assess extreme water-level events that have a low probability of occurring such as those with return periods greater than the 100-yr event. This is especially important because the Federal Emergency Management Agency and the U.S. Army Corps of Engineers highly recommend that critical infrastructure or structures protecting critical infrastructure (e.g., levees and floodwalls) are built to withstand the 500-yr event plus freeboard \cite{fema-543, US_Army_Corps_of_Engineers2018-nv}. Critical infrastructure are facilities that provide essential services to the community, protect the community, and are intended to remain open during and after major disasters. Such facilities include health care facilities, schools and higher education buildings, facilities storing hazardous material, and emergency response facilities (e.g., fire stations, police stations, and emergency operation centers) \cite{fema-543}. For example, in collaboration with the City of Norfolk’s resilience initiative and the U.S. Army Corps of Engineers, the Sentara Norfolk General Hospital (a level one trauma center located in the current 100-yr and 500-yr floodplains) is currently implementing measures to protect the hospital against sea-level rise and storm surge in its five-year project (construction to be completed in 2020) to expand and modernize the hospital, in addition to a proposed storm surge barrier that would help protect the hospital and other nearby critical infrastructure \cite{Norfolk-2015-rs, gauding_2016, US_Army_Corps_of_Engineers2018-nv}.

Despite a very low probability of occurring, these events do occur (Figure~\ref{figfour}A). Over the course of 70 years (the potential useful lifetime of a building), a 500-yr event has an 18\% probability of occurring \cite{fema-543}. When these events occur, they are often considered high-impact disasters. Consider, for example, the storm surge of Hurricane Sandy, a roughly 400-yr event at the Battery in New York City \cite{Lin2016-rm}. More importantly, consider the lesser known (in modern history), but violent Norfolk-Long Island Hurricane of 1821 and Storm of 1749. The Norfolk-Long Island Hurricane made landfall on September 3rd, 1821, hitting Norfolk, VA, among other major cities along the Mid-Atlantic and East Coast \cite{Ludlum1963-tg, Linkin2014-ih}. The hurricane is estimated to have caused a storm surge of roughly 10 ft in some areas of the Virginia coastline \cite{Ludlum1963-tg, Linkin2014-ih} (Figure~\ref{figfour}A). This is an approximately 285-yr storm surge event (approximated using a method to calculate the median probability return period; see Methods). Based on historical records, we can reasonably constrain the uncertainty bounds between the 279- and 306-yr storm surge event as a larger storm surge event may have occurred during a hurricane in 1825 and less documentation exists for storm surges prior to 1806 \cite{Ludlum1963-tg}. A study by the reinsurance company Swiss Re estimates that the Norfolk-Long Island Hurricane of 1821 would cause 50\% more damage than Hurricane Sandy and more than 100 billion U.S. dollars in damages if it were to occur today \cite{Ludlum1963-tg, Linkin2014-ih}. The Storm of 1749 (a hurricane) hit the Mid-Atlantic coast during October of 1749 causing roughly 30 thousand pounds in damages in Norfolk at that time \cite{Ludlum1963-tg, Hampton-2017-hm}. During this storm, a 15 ft storm surge and subsequent flooding of the lower Chesapeake Bay area was reported \cite{Ludlum1963-tg, Hampton-2017-hm} (Figure~\ref{figfour}A). Moreover this storm was largely responsible for forming Willoughby Spit, 2 mile long and 1/4 mile wide peninsula of land at Sewell's point \cite{Hampton-2017-hm}. The storm surge of the Storm of 1749 is approximately a 389-yr storm surge event (approximated using a method to calculate the median probability return period; see Methods). Despite the potential high impact of low probability events and their importance in resilience planning, our model and Wong~\cite{Wong-2018} are the only models in this analysis to resolve return periods past the 100-yr event (Figure~\ref{figfour}A).

It is important also to note that storm surges may not be stationary (i.e., the statistics are not constant over time). While decision makers typically use stationary flood hazard information, neglecting non-stationary information can result in a low bias (Figure~\ref{figfour}B). In the year 2065, all non-stationary 100-yr storm surge values are greater than the storm surge values based on stationary models with a difference of up to a foot (with the exception of the non-stationary case based on the NAO index covariate time series; Figure~\ref{figfour}B). To put this into context, a difference of less than a foot in storm surge values can be the difference of millions of dollars in potential damages to the Norfolk area~\cite{Furgo-2016}. Despite the importance of considering the non-stationary behavior of storm surge projections, it is still an area of active research. In particular, there is an active debate on how best to account for and constrain potential non-stationary coastal surge behavior \cite{Wahl-2016, Wong-2018}. For example, a recent study by Wong~\cite{Wong-2018} analyzes how different climate variable time series (e.g., temperature, sea level, NAO index, and time) impact non-stationary storm surge values. Although non-stationary storm surge is an area of active research, non-stationary flood hazard information is available and can be used by decision makers.

\subsection*{Accounting for uncertainty in projected future flood hazards}
\paragraph{}

\begin{figure}[ht]
\centering
\includegraphics{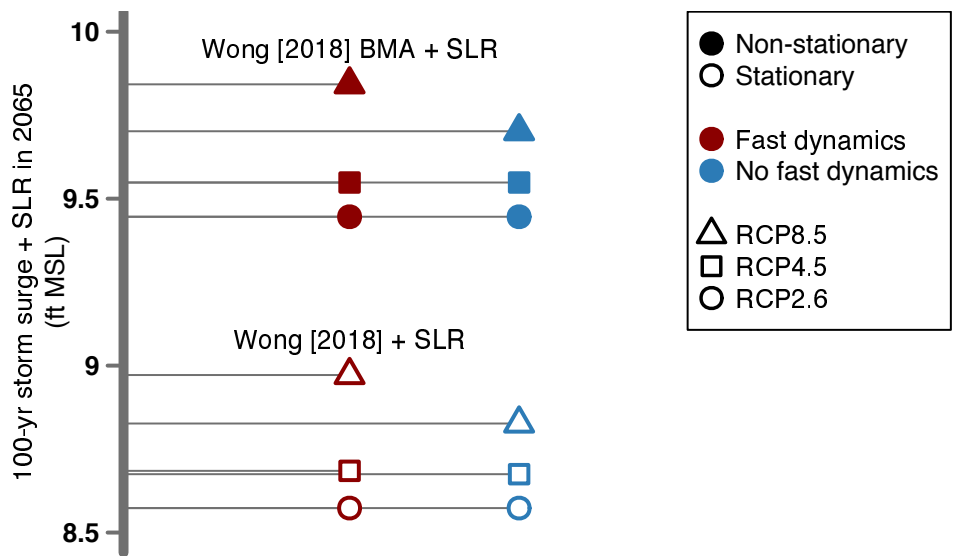}
\caption{Comparison of the 100-yr combined storm surge and sea-level rise in the year 2065 accounting for uncertainty. Note that larger values cannot be excluded (for example in the event of compound flooding).}
\label{figfive}
\end{figure}

Combining probabilistic projections of local SLR with storm surge analysis more accurately assesses flood risk (details in Methods; Figure~\ref{figfive}). The combined SLR and storm surge projections increase estimates of future flood risk (see also the discussion in \cite{Ruckert2017-an}; Figure~\ref{figfour}B versus Figure~\ref{figfive}). Despite evaluating multiple studies, only the SLR projections from Kopp et al.~\cite{Kopp2014-ri}, Sweet et al.~\cite{Sweet2017-uh}, Wong and Keller~\cite{Wong2017-kn}, and Rasmussen et al.~\cite{Rasmussen2018-uy} provide enough information (i.e., full probability distributions) to account for interactions between uncertainties. Likewise our study and Wong~\cite{Wong-2018} are the only studies in this analysis that provide full probability distributions of storm surge estimates. To reduce complexity, we show four cases of the combined SLR and storm surge projections accounting for uncertainty. These cases depict how fast dynamics and non-stationarity impact storm surge estimates (Figure~\ref{figfive}). In particular, there is a roughly 0.87 ft difference between stationary and non-stationary combined storm surge and SLR estimates. Additionally, combined storm surge and SLR estimates increase as ice-cliff and ice shelf feedback processes are triggered in the year 2065, which are triggered in high emissions scenarios (i.e., a roughly 0.14 ft increase when triggered using RCP8.5; Figure~\ref{figfive}). Moreover, these few cases highlight the complexity of estimating a single point when there are multiple uncertainties to consider.

\section*{Discussion and caveats}
\paragraph{}

The choice of methods and assumptions used in a flood hazard study can impact the design of flood risk management strategies. These choices can limit the amount of information available to a vulnerable community interested in coastal preparedness planning. For instance, a community concerned about ice sheet feedback processes, individual components comprising SLR, long timescales, uncertainty, and the 500-yr storm surge event is limited in the considered sample of studies to the results in Kopp et al.~\cite{Kopp2017-fz} and our extended projections of Kopp et al.~\cite{Wong2017-kn} FD for SLR projections, and the method outlined in this study and the results in Wong~\cite{Wong-2018} for storm surge projections (Figure~\ref{figfour}). This lack of information reduces the range of choices in the decision-making process and hence could lead to poor outcomes. The potential consequences of having insufficient data are 1) cities are unprepared for extreme events (i.e., like the 1821 Norfolk-Long Island Hurricane and/or abrupt changes in ice-sheet dynamics; Figure~\ref{figfour}A and Figure~\ref{figfive}) or 2) cities over-invest in protection measures.

We analyze and synthesize multiple future SLR scenarios and storm-surge characterizations generated from different approaches. Our findings help to understand and quantify the sources and effects of the deep uncertainty surrounding coastal flood hazard projections. Our results highlight some of the current limitations of coastal flood hazard characterizations when used to inform the design of strategies to manage flood risks. Coastal flood hazard projections diverge considerably across decision-relevant timescales based on the adopted methods and assumptions. Relying on a single flood hazard projection can hence be interpreted as making a deeply uncertain gamble. This deep uncertainty stems, for example, from the difficulties associated with calibrating the model parameters and the divergent expert assessments. The models need to resolve the complex responses of glaciers and ice sheets where understanding the physics is important, but often challenging. As a consequence, modeled results and expert elicitation results can be biased and/or overconfident. One approach to inform decisions in the face of these deep and dynamic uncertainties is to adopt decision-making approaches that allow for the adaptation of decisions over time to meet changing circumstances, respond to abrupt changes, promote continual learning, and revisit coastal flood hazard projections in the future when necessary knowledge about uncertain physics is, hopefully, better understood \cite{Haasnoot-2013, Garner-2018}. One such approach is referred to as Dynamic Adaptive Policy Pathways. This approach searches for an adaptive plan that can deal with changing conditions and “supports the exploration of a variety of relevant uncertainties in a dynamic way, connects short-term targets and long-term goals, and identifies short-term actions while keeping options open for the future” \cite{Haasnoot-2013}. Consider, as an example, the flexible adaptation pathways approach taken by the City of New York in their climate action strategy, which allows the city plan to adapt over time \cite{Rosenzweig-2011, Rosenzweig-2014}. Due to adopting this approach, the city was able to revisit plans and respond in the aftermath of Hurricane Sandy \cite{Rosenzweig-2014}.

Despite our focus on coastal flood hazard assessment and coastal preparedness planning, we neglect the interaction between different hazard drivers beyond the interaction of uncertainties between SLR and storm surge. Specifically, we neglect to address the issue of compound flooding. Compound flooding refers to flooding caused by a combination of multiple drivers and/or hazards such as SLR, storm tides and waves, precipitation, or river discharge that lead to societal or environmental impacts \cite{Sadegh-2018, Zscheischler-2018}. For example, in 2017, compound flooding occurred in Jacksonville, FL and the greater Houston area because of the combination of storm surge and high discharge of the St. John’s river during Hurricane Irma and the combination of extreme precipitation and long-lasting storm surge during Hurricane Harvey, respectively \cite{Sadegh-2018, Zscheischler-2018}. We chose to neglect compound flooding in our assessment to provide a transparent analysis on what causes SLR and storm surge projections to diverge and how this divergence impacts coastal planning. This implies that larger values and uncertainties cannot be excluded when considering other hazard drivers that interact with SLR. Due to the importance of this issue, it is necessary to consider and analyze compound flooding in coastal risk assessments, decision-making, and future research. 

\section*{Conclusions}
\paragraph{}
Coastal communities rely on flood hazard projections to design risk management strategies. Studies evaluating future flood hazards often provide only a limited description of the deep uncertainties surrounding these projections and diverge in projections. Using Norfolk, VA, as an example, we show how the lack of information (i.e., extreme cases, non-stationarity, and ice sheet feedbacks) can lead to surprises. We highlight the importance of estimating the different components of SLR and accounting for Antarctic ice sheet fast dynamics, especially when ice sheet contributions play a greater role in SLR at the end of this century and beyond \cite{DeConto2016-dp}. Even though the considered studies that produce plausible scenarios provide a broad range of future conditions, they  do not produce probability distributions. Without probabilistic distributions, we could not evaluate the combination of SLR and storm surge while accounting for interactions with uncertainties nor could we evaluate differences in the 500-yr return period between multiple studies. Although we evaluate flood hazard projections for Norfolk, our conclusions are transferable to many regions. Improving the representation of ice-sheet feedback processes, decision-relevant timescales, extreme events, non-stationarities, and known unknowns can improve risk assessments and decision-making.

\section*{Methods}
\subsection*{Comparing data}
\paragraph{}
We identify key existing studies relevant to the case-study location and then identify the background assumptions and methods for each study. Each study presents projections in different units of measurement, relative to different datums, start at different years, and do not always incorporate local rates of subsidence. For instance, Tebaldi et al.~\cite{Tebaldi2012-ou} presents storm surge projections in meters above the mean high water datum, whereas Zervas~\cite{Zervas2013-sv} presents storm surge projections in meters above the mean higher high water datum. Following a detailed review of these studies, we modify the projections to the same baseline conditions for comparability. For comparability and consistency across all SLR projections, we modify the scenarios and projections relative to the local mean sea level, update the start year to 2000, present projections in feet, and incorporate local subsidence in all projections. For storm surge, projections are in feet above the local mean sea level for the current NTDE (1983-2001). The NTDE represents the period of time used to define the tidal datum (e.g., mean high water and local mean sea level) \cite{Flick2013-ef}.

\subsection*{Extending and localizing sea-level rise projections}
\paragraph{}
Unlike Wong and Keller~\cite{Wong2017-kn}, we project the BRICK model to the year 2200 using the RCP2.6, 4.5, and 8.5 radiative forcing scenarios \cite{Meinshausen2011}. Wong and Keller~\cite{Wong2017-kn} only project the model to the year 2100 because the Greenland ice sheet, and glaciers and ice cap models do not account for what happens when the ice mass completely melts. In short, the models do not simulate the possibility of a net gain of ice. Hence, once the ice mass is at zero, there is no regrowth of ice. Running a simple diagnostic test, we test the reliability of our projections out to 2200 and find that only the ice mass from the glaciers completely melts in some high emissions scenarios with a lower initial ice mass (Supplementary Figure 1) \cite{Wong2017-th}. We then downscale global sea-level projections using sea-level fingerprints from Slangen et al.~\cite{Slangen2014-mq} and localize sea-level projections by supplying the model with the coordinates of the Sewell's Point tide gauge. These projections do not incorporate local subsidence; therefore, we add in the long-term, local, non-climatic sea-level change projections from Kopp et al.~\cite{Kopp2014-ri}.

With the exception of Sweet et al.~\cite{Sweet2017-uh}, we downscale each of the sea-level rise studies based on the method of providing plausible scenarios (i.e., \cite{US_Army_Corps_of_Engineers2011-fm, US_Army_Corps_of_Engineers2013-oh, US_Army_Corps_of_Engineers2014-jt, Parris2012-oz, Hall2016-fh}). Following the U.S. Army Corps of Engineers~\cite{US_Army_Corps_of_Engineers2011-fm, US_Army_Corps_of_Engineers2013-oh, US_Army_Corps_of_Engineers2014-jt} approach, we downscale these scenarios to the local level using the local mean sea-level trend of 4.44 mm/yr \cite{Zervas2013-qf} as the rate of SLR in the National Research Council~\cite{National_Research_Council1987-oj} global sea-level model. The local mean sea-level trend at the Sewell's Point tide gauge accounts for local and regional vertical land movement, coastal environmental processes, and ocean dynamics \cite{US_Army_Corps_of_Engineers2011-fm, US_Army_Corps_of_Engineers2013-oh, US_Army_Corps_of_Engineers2014-jt, Zervas2013-qf}.

\subsection*{Storm surge observations and projections}
\paragraph{}
To generate historical observations with associated return periods (Figure~\ref{figfour}a and Supplementary Figure 2), we use hourly records of observed water levels from the Sewell’s Point tide gauge \cite{National_Oceanic_and_Atmospheric_Administration2013-ly}. The observed water levels are relative to the MSL datum of the NTDE. The record we use is 88 years in length from 1928 to 2015. These observations contain the longer-term signal (SLR), which masks the effects of day-to-day weather, tides, and seasons. Following previous work (e.g., \cite{Oddo-pc, Ruckert2017-an}), we subtract the annual means from the record to approximately remove the SLR trend. We then approximate the annual block maxima by grouping the values into non-overlapping annual observation periods (Supplementary Figure 2a). To calculate the return period associated with the annual block maxima (Supplementary Figure 2b), we use a numerical median probability return period method \cite{Jenkinson1977-af, Folland2002-fa, Makkonen2008-la}. We use this method to reduce plotting biases by calculating the median probability of a return period instead of the mean \cite{Jenkinson1977-af, Folland2002-fa, Makkonen2008-la}. This method calculates the median probability of a return period for an annual block maxima by estimating the binomial distribution that places the ranked event at the median of the distribution \cite{Jenkinson1977-af, Folland2002-fa, Makkonen2008-la}.

Similar to the approach in Oddo et al.~\cite{Oddo-pc}, our model uses hourly tide gauge data from 1926 to 2016 and a Bayesian calibration approach to fit an ensemble of stationary GEV distributions for the storm surge projections at the Sewell's Point tide gauge. To set up the GEV analysis, we first subtract the annual means from the tide gauge record followed by calculating the annual block maxima from the detrended record. Using the detrended record, we calculate a maximum likelihood estimate for the GEV distribution parameters using the extRemes R package \cite{Gilleland2013-rj, R_Core_Team2016-yv}. The resulting estimates act as the starting point for a 500,000-iteration Markov chain Monte Carlo simulation of the GEV parameters. We discard the first 50,000 iterations of each chain to remove the effects of starting values.

\subsection*{Combining sea-level and storm surge projections}
\paragraph{}
Accounting for the interactions between SLR and storm surge is a crucial step in the assessment of coastal flood vulnerability. Specifically, it is necessary to account for the uncertainties surrounding SLR (see, for example, the discussion in \cite{Ruckert2017-an}). Similar to the approach in Ruckert et al.~\cite{Ruckert2017-an}, we account for uncertainty in both SLR and storm surge by combining the distributions. Following this approach, we combine the Wong and Keller~\cite{Wong2017-kn} no FD SLR distribution for the year 2065 with the ensemble of stationary storm surge values obtained in Wong~\cite{Wong-2018}. Because the distributions differ in ensemble size, we draw an ensemble size of 10,000 simulations of storm surge (each simulation projecting out to the 1000-yr return period) from the full distribution to correspond to the SLR ensemble size. We then approximate the 100-yr storm surge following the steps outlined in Ruckert et al.~\cite{Ruckert2017-an} (see paper for details). Following the same procedure, we also combine 1) Wong and Keller~\cite{Wong2017-kn} FD SLR distribution with the stationary storm surge in Wong~\cite{Wong-2018}, 2) Wong and Keller~\cite{Wong2017-kn} no FD SLR distribution with the non-stationary BMA storm surge in Wong~\cite{Wong-2018}, and 3) Wong and Keller~\cite{Wong2017-kn} FD SLR distribution with the non-stationary BMA storm surge in Wong~\cite{Wong-2018}. It is also possible to use the sea-level distributions from Sweet et al.~\cite{Sweet2017-uh}, Kopp et al.~\cite{Kopp2014-ri}, and Rasmussen et al.~\cite{Rasmussen2018-uy}, as well as the other non-stationary storm surge ensembles presented in Wong~\cite{Wong-2018}; however, we choose to show the four cases stated above to make clear the differences between SLR with and without fast dynamics and stationary versus non-stationary storm surge.

\section*{Acknowledgements (not compulsory)}

We thank Robert Nicholas, Nancy Tuana, Irene Schaperdoth, Ben Lee, Francisco Tutella, and Randy Miller for their valuable inputs. We also thank Claudia Tebaldi and Tony Wong for sharing data and for their valuable inputs. Additionally, we thank K. Joel Roop-Eckart for sharing his function approximating median probability return periods for observations. This work was supported by  the National Oceanic and Atmospheric Administration (NOAA) Mid-Atlantic Regional Integrated Sciences and Assessments (MARISA) program under NOAA grant NA16OAR4310179 and the Penn State Center for Climate Risk Management. We are not aware of any real or perceived conflicts of interest for any authors. Any conclusions or recommendations expressed in this material are those of the authors and do not necessarily reflect the views of the funding agencies. Any errors and opinions are, of course, those of the authors.

\section*{Author contributions statement}

K.L.R drafted the main manuscript text, prepared the figures, and conducted the analysis. V.S. performed the Bayesian calibration for the storm surge analysis. K.K. initiated the study. All authors discussed the results and edited the manuscript.

\section*{Additional information}
\textbf{Supplementary information} accompanies this paper.

\noindent\textbf{Competing interests:} The authors declare no competing interests.

\noindent\textbf{Data availability:} All code, data, and output are available on Data Commons (\href{https://doi.org/10.26208/z5e5-kh11}{DOI: 10.26208/z5e5-kh11}) and \url{http://www.github.com/scrim-network/local-coastal-flood-risk} under the GNU general public open-source license. Data and analysis codes for our storm surge model are located at \url{http://www.github.com/vsrikrish/SPSLAM}. The results, data, software tools, and other resources related to this work are provided as-is without warranty of any kind, expressed or implied. In no event shall the authors or copyright holders be liable for any claim, damages or other liability in connection with the use of these resources.

\bibliography{main}

\bibliographystyle{unsrt} 

\end{document}